\documentclass[reprint,aps,prb,amsmath,amssymb,nobibnotes,superscriptaddress,10pt]{revtex4-2}
\usepackage[T1]{fontenc}
\usepackage{textcomp}
\usepackage{txfonts}
\usepackage{bm}

\usepackage[normalem]{ulem}
\usepackage{graphicx}
\usepackage{subcaption}
\usepackage{csquotes}
\usepackage[english]{babel}
\usepackage[letterpaper,top=2cm,bottom=2cm,left=3cm,right=3cm,marginparwidth=1.75cm]{geometry}
\usepackage{amsmath}
\usepackage{graphicx}
\usepackage[colorlinks=true, allcolors=blue]{hyperref}
\usepackage{booktabs}
\usepackage{multirow}

\usepackage{enumitem}
\usepackage[noabbrev]{cleveref}

\usepackage{caption}
\usepackage{listings}
\usepackage[usenames,dvipsnames]{xcolor}

\definecolor{codegray}{rgb}{0.5,0.5,0.5}
\definecolor{backcolour}{rgb}{0.95,0.95,0.92}
\definecolor{colorcomments}{rgb}{0.25,0.5,0.5}
\definecolor{colorkeywords}{rgb}{0,0.6,0}
\definecolor{colorstring}{HTML}{BA2121}
\definecolor{colornumber}{HTML}{FF0000}

\lstdefinestyle{mystyle}{
    backgroundcolor=\color{backcolour},   
    commentstyle=\color{colorcomments},
    keywordstyle=\color{colorkeywords}\bfseries,
    numberstyle=\tiny\color{codegray},
    stringstyle=\color{colorstring},
    basicstyle=\ttfamily,
    breakatwhitespace=true,         
    breaklines=true,                 
    captionpos=b,                    
    keepspaces=true,                 
    numbers=left,                    
    numbersep=5pt,                  
    showspaces=false,                
    showstringspaces=false,
    showtabs=false,                  
    tabsize=2,
    framexleftmargin=5mm, 
    framexrightmargin=5mm, 
    frame=tb, 
    framerule=1pt, 
    framextopmargin=3mm,
    framexbottommargin=3mm
}

\crefname{lstlisting}{code}{code}
\Crefname{lstlisting}{Code}{Code}

\usepackage{tcolorbox}
\usepackage{tabularx}
\usepackage{array}
\usepackage{colortbl}
\tcbuselibrary{skins}

\newcolumntype{Y}{>{\raggedleft\arraybackslash}X}

\tcbset{tab2/.style={standard,fontupper=\normalsize\sffamily,
colback=backcolour}}

\begin{document}

\title{Comparative Study of Quantum Transpilers: Evaluating the Performance of qiskit-braket-provider, qBraid-SDK, and Pytket Extensions}

\author{Mohamed Messaoud \surname{Louamri}}
\thanks{© 2024 IEEE.  Personal use of this material is permitted.  Permission from IEEE must be obtained for all other uses, in any current or future media, including reprinting/republishing this material for advertising or promotional purposes, creating new collective works, for resale or redistribution to servers or lists, or reuse of any copyrighted component of this work in other works. \\
Published version: \, \url{https://doi.org/10.1109/IC3IT63743.2024.10869429}}
\affiliation{Constantine Quantum Technologies, \\University Constantine 1 Fr\`{e}res Mentouri, Ain El Bey Road, Constantine, 25017, Algeria}
\affiliation{qBraid Co., 111 S Wacker Dr., Chicago, IL 60606, USA}
\affiliation{Theoretical Physics Laboratory, \\University of Science and Technology
Houari Boumediene, BP 32 Bab Ezzouar, Algiers, 16111, Algeria}

\author{Nacer Eddine \surname{Belaloui}}  
\affiliation{Constantine Quantum Technologies, \\University Constantine 1 Fr\`{e}res Mentouri, Ain El Bey Road, Constantine, 25017, Algeria}
\affiliation{qBraid Co., 111 S Wacker Dr., Chicago, IL 60606, USA}
\affiliation{Laboratoire de Physique Math\`{e}matique et Subatomique, \\University Constantine 1 Fr\`{e}res Mentouri, Ain El Bey Road, Constantine, 25017, Algeria}
\author{Abdellah \surname{Tounsi}}  
\affiliation{Constantine Quantum Technologies, \\University Constantine 1 Fr\`{e}res Mentouri, Ain El Bey Road, Constantine, 25017, Algeria}
\affiliation{qBraid Co., 111 S Wacker Dr., Chicago, IL 60606, USA}
\affiliation{Laboratoire de Physique Math\`{e}matique et Subatomique, \\University Constantine 1 Fr\`{e}res Mentouri, Ain El Bey Road, Constantine, 25017, Algeria}
\author{Mohamed Taha \surname{Rouabah}}
\email[Corresponding author: ]{taha.rouabah@cqtech.org}
\affiliation{Constantine Quantum Technologies, \\University Constantine 1 Fr\`{e}res Mentouri, Ain El Bey Road, Constantine, 25017, Algeria}
\affiliation{qBraid Co., 111 S Wacker Dr., Chicago, IL 60606, USA}
\affiliation{Laboratoire de Physique Math\`{e}matique et Subatomique, \\University Constantine 1 Fr\`{e}res Mentouri, Ain El Bey Road, Constantine, 25017, Algeria}

\begin{abstract}
In this study, we present a comprehensive evaluation of popular SDK-to-SDK quantum transpilers (that is transpilers that takes a quantum circuit from an initial SDK and output a quantum circuit in another SDK), focusing on critical metrics such as correctness, failure rate, and transpilation time. To ensure unbiased evaluation and accommodate diverse quantum computing scenarios, we developed two dedicated tools: RandomQC, for generating random quantum circuits across various types (pure random, VQE-like, and SDK-specific circuits), and Benchmarq, to streamline the benchmarking process. Using these tools, we benchmarked prominent quantum transpilers as of February 2024. Our results highlight the superior performance of the qiskit-braket-provider, a specialized transpiler from Qiskit to Braket, achieving a remarkably low failure rate of 0.2\%. The qBraid-SDK, offering generalized transpilation across multiple SDKs, demonstrated robust but slower performance. The pytket extensions, while fast, faced limitations with complex circuits due to their one-to-one transpilation approach. In particular, the exceptional performance of the qiskit-bracket-provider stems not only from its specialization but also from its architecture, which combines one-to-one transpilation with gate decomposition for unsupported gates, enhancing both speed and capability. This study aims to provide practical guidelines to users of SDK-to-SDK quantum transpilers and guidance to developers for improving the design and development of future tools.
\end{abstract}

\maketitle

\section{Introduction}

Quantum computing holds the promise to transform our computational power and revolutionize our problem-solving capabilities\cite{Nielsen_Chuang_2010,Arute2019,Madsen2022}. In a relentless race toward practical implementations, companies strive to construct large, cost-effective, and robust quantum computing platforms. To this end, technologies such as superconductors\cite{Kjaergaard2020}, trapped ions\cite{Monroe2013}, neutral atoms\cite{Henriet2020}, photonics\cite{OBrien2009}, etc. are harnessed, each presenting unique advantages and challenges.
Amidst the hardware race, a parallel effort unfolds in software development. Abstraction of physics inherent to diverse hardware platforms and streamlining quantum software development necessitate the creation of programming languages and software development kits (SDKs). 
Various SDKs have emerged to facilitate quantum programming such as IBM's Qiskit\cite{qiskit}, Google's Cirq\cite{cirq}, Amazon's Braket\cite{braket}, Rigetti's Quil\cite{quil}, Xanadu's Pennylane\cite{pennylane}, and Quantinuum's Pytket\cite{pytket}, among others. As the field is still in its nascent stage, each company's focal points, principles, and applications are different, leading to each SDK carrying its own philosophy, capabilities, and limitations.
Recognizing the impracticality of rewriting code for each SDK and for each quantum computer. SDK-to-SDK quantum transpilers have emerged as a crucial bridge in the quantum computing ecosystem. These transpilers enable users to craft quantum circuits using their SDK of choice and seamlessly convert them into equivalent circuits compatible with different SDKs. This allows users to exploit the unique capabilities of each SDK and execute circuits on multiple quantum computers without rewriting them.
%

This study aims to provide practical guidelines to users of SDK-to-SDK quantum transpilers and guidance to developers for improving the design and development of future tools. At the time of writing this article (Q1 2024), we are not aware of any similar study that comprehensively evaluates SDK-to-SDK quantum transpilers.

This paper is structured as follows. In the subsequent section, we will elaborate on the methodology employed to benchmark various quantum transpilers. Section \ref{sec:res} will then showcase our results, and in Section \ref{sec:conc}, we will provide a concise conclusion along with our recommendations.
\section{Methodology} \label{sec:met}
To achieve the objectives outlined in the introduction, a systematic methodology was employed to benchmark various quantum transpilers. The setup involved the generation of random quantum circuits spanning various sizes and depths. Some of which were designed to reflect real-world quantum algorithms and computational problems. The diversity in circuit characteristics allows us to assess the transpilers' capabilities for different quantum computing scenarios.
\subsection{Random Circuit Generation}
To generate our random quantum circuits, multiple choices were considered but generators like \texttt{cirq.testing.random\_circuit} and \texttt{qiskit.circuit.random.random\_circuit} yield circuits tailored to their respective SDKs. On the other hand,  generators like \texttt{qbraid.interface.random\_circuit} can produce random circuits for multiple SDKs, but they employ their transpilers in the process. Using them could potentially introduce biases into our evaluation, and for this reason, it was avoided. Additionally, existing methods generate circuits that are purely random without a specific shape or architecture.
To overcome the aforementioned limitations, we developed a dedicated tool labeled \texttt{RandomQC}. At the core of this package is the \texttt{RandomCircuit} abstract base class, which incorporates two attributes: 
\begin{itemize}
    \item \texttt{name}: represents a descriptive name of the random circuits to be generated, serving as a tag for evaluating transpiler performance based on circuit types.
    \item  \texttt{qasm2}: represents the QASM2 representation of the generated circuit.
\end{itemize}
The methods of this class include:
\begin{itemize}
    \item \texttt{\_generate}: an internal abstract method for generating the quantum circuit as a qasm2 string with a specific architecture.
    \item \texttt{to}: a method to convert the circuit from qasm2 to the circuit specific to the target SDK. To avoid introducing bias during this conversion, we have used the utility functions provided by the target SDK to create circuits from QASM2 strings (e.g: for qiskit, we have used \texttt{QuantumCircuit.from\_qasm\_str}). 
\end{itemize}
 
This class is designed to be inherited, necessitating the implementation of abstract methods by the child class to define the architecture of the random quantum circuit. In addition to these attributes and methods, users have the flexibility to add their own attributes and methods as needed.
As some SDKs eliminate idle qubits from their circuits, which can result in a mismatch in the size of the matrix representation of the circuits when comparing the initial quantum circuit to its transpiled counterpart. We chose to ensure the absence of idle qubits in the initial quantum circuit (those generated by \texttt{RandomQC} or by its plugins). Consequently, the minimum number of gates in our randomly generated circuits is set to be equal to or greater than the number of qubits, and the package ensures that there is at least one gate per qubit. This condition serves to simplify the benchmarking process and does not introduce any bias.

As examples of child classes derived from the \texttt{RandomCircuit} class, we have the \texttt{PureRandom} class and the \texttt{VQE} class. The \texttt{PureRandom} class is utilized to generate random circuits without a specific architecture. The class randomly draws gates from a predefined set, defined in \texttt{qelib1.inc} library. The set includes single-qubit gates: \texttt{x}, \texttt{y}, \texttt{z}, \texttt{h}, \texttt{s}, \texttt{sdg}, \texttt{t}, \texttt{tdg}, \texttt{sx}, \texttt{sxdg}, \texttt{p}, \texttt{rx}, \texttt{ry}, \texttt{rz}, as well as two-qubit gates: \texttt{cx}, \texttt{cy}, \texttt{cz}, \texttt{ch}, \texttt{swap}, \texttt{crx}, \texttt{cry}, \texttt{crz}, \texttt{cp}, \texttt{rxx}, and \texttt{rzz}.  The class is instantiated with two parameters: 
\begin{itemize}
    \item \texttt{nb\_qubits}: denotes the number of qubits in the circuit.
    \item \texttt{nb\_gates}: specifies the number of gates in the circuit such that $\texttt{nb\_qubits} \leq \texttt{nb\_gates}$.
\end{itemize}
By default, \texttt{nb\_qubits} is randomly picked within the range [2, 10] and \texttt{nb\_gates} is randomly picked within the range [2, 100]. For convenience, a static method, \texttt{range\_gates}, has been incorporated. This method takes the following parameters:
\begin{itemize}
    \item \texttt{nb\_qubits}: the number of qubits in the circuit.
    \item \texttt{min\_nb\_gates}: the minimum number of gates in the generated circuits. The condition $\texttt{nb\_qubits} \leq \texttt{min\_nb\_gates}$ should be satisfied.
    \item \texttt{max\_nb\_gates}: the maximum number of gates in the generated circuits.
    \item \texttt{gate\_step=1}: an optional parameter defining the step size between the gate counts in the generated list.
\end{itemize}
The  \texttt{range\_gates} method generates a list of random circuits starting from \texttt{min\_nb\_gates} up to \texttt{max\_nb\_gates} (not included), with a step defined by \texttt{gate\_step}. This can be beneficial for assessing how certain transpiler metrics scale with the circuit gate count, for example, to study transpilation time as a function of gatecount. An example of utilizing the \texttt{PureRandom} class is provided in \cref{ex_pure_random}. In this example, ``qc1" is created with 2 qubits and 4 gates, resulting in a random circuit with the specified parameters. Subsequently, ``qc2" generates a random circuit with parameters randomly chosen such that the number of qubits falls within the range [2, 10], and the number of gates falls within the range [2, 100], the variable ``qcs" is assigned a list of random circuits, each containing 2 qubits, with the number of gates ranging from 2 to 100. The \texttt{to} method is used to convert \texttt{qc1} and the 3\textsuperscript{rd} element of \texttt{qcs} to a pytket circuit, the results are stored in \texttt{tket1} and \texttt{tket2}, respectively.

\begin{lstlisting}[language=Python,caption={Example of Using the \texttt{PureRandom} Class},label={ex_pure_random},float=*]
>>> from randomqc import PureRandom

>>> qc1 = PureRandom(nb_qubits=2, nb_gates=4)
>>> qc2 = PureRandom()
>>> qcs = PureRandom.range_gates(2, 2, 101)
>>> tket1 = qc1.to('pytket')
>>> tket1
[Sdg q[0]; T q[1]; CZ q[0], q[1]; CY q[1], q[0]; ]

>>> tket2 = qcs[3].to('pytket') 
>>> tket2
[T q[0]; X q[1]; U1(1.46706) q[1]; CRy(0.374142) q[0], q[1]; ]
\end{lstlisting}

On the other hand, the \texttt{VQE} class generates random quantum circuits following the architecture of Variational Quantum Eigensolver (VQE) circuits, comprising a parameterized ansatz followed by a Pauli string. The \texttt{VQE} class accepts two arguments, \texttt{nb\_qubits} and \texttt{ansatz}; if left empty, the arguments will be randomly selected. In the version presented in this study, \texttt{RandomQC} implements only one type of ansatz, the Hardware Efficient Ansatz (HEA). This ansatz is also randomly generated and accepts four parameters: 
\begin{itemize}
    \item \texttt{nb\_qubits}: which should align with the VQE's circuit number of qubits.
    \item  \texttt{nb\_su2\_gates}: specifying the number of SU2 gates in the rotation layer.
    \item \texttt{entanglement}: specifying the entanglement structure (either ``linear" or ``circular").
    \item \texttt{reps}: specifying the number of repetitions of the rotation+entanglement layer.
    
\end{itemize}
Arguments for the ansatz can be directly passed through the \texttt{VQE} constructor. An example of utilizing the \texttt{VQE} class is provided in \cref{ex_vqe}. In this example, ``qc3" is generated with randomly chosen arguments such that the number of qubits falls within the range [2, 10], and the ansatz is selected as HEA with random arguments. That is, the number of SU2 gates is randomly chosen within the range [1, 3], the type of entanglement is randomly selected as either ``linear" or ``circular," and the repetitions are randomly chosen within the range [1, 4]. The circuit ``qc4" is generated with the following parameters: the number of qubits is set to 2, and for the ansatz, the number of SU2 gates is set to 1, the entanglement is chosen as ``linear," and the repetitions are set to 4. The circuits are then converted to pytket circuits and stored in \texttt{tket3} and \texttt{tket4}, respectively.
%
\begin{lstlisting}[language=Python,caption={Example of Using the \texttt{VQE} class},label={ex_vqe},float=*]
>>> from randomqc import VQE

>>> qc3 = VQE()
>>> tket3 = qc3.to('pytket')
>>> tket3
[Rx(0.501217) q[0]; Rx(1.88348) q[1]; Ry(1.63825) q[0]; Ry(1.90287) q[1]; CX q[0], q[1]; Rx(1.68886) q[0]; Rx(0.516951) q[1]; Ry(0.82392) q[0]; Ry(0.584921) q[1]; CX q[0], q[1]; Rx(0.73289) q[0]; Rx(0.46376) q[1]; Ry(0.686261) q[0]; Ry(0.292793) q[1]; H q[0]; ]

>>> qc4 = VQE(nb_qubits=2, nb_su2_gates=1, entanglement='linear', reps=3)
>>> tket4 = qc4.to('pytket')
>>> tket4
[Rx(1.17985) q[0]; Rx(1.59042) q[1]; CX q[0], q[1]; Rx(0.986453) q[0]; Rx(0.260294) q[1]; CX q[0], q[1]; Rx(0.781044) q[0]; Rx(1.19931) q[1]; CX q[0], q[1]; Rx(1.46526) q[0]; Rx(1.31492) q[1]; ]
\end{lstlisting}
%
Additionally, to evaluate the transpilation performance of specific gates and circuit generation routines defined by various SDKs, we developed SDK-specific plugins for the \texttt{RandomQC} package. These plugins extend the capabilities of \texttt{RandomQC} to generate circuits incorporating SDK-specific elements. An example of such plugins is \texttt{randomqc\_qiskit}, which generates random Qiskit \texttt{QuantumCircuit}s utilizing Qiskit's specific functions. The primary routine of \texttt{randomqc\_qiskit} is \texttt{QiskitGates.get\_all()}, which returns a dictionary of all Qiskit circuits the plugin can generate. An example of \texttt{randomqc\_qiskit} is found in the listing. In it, a total of 102 random circuits are generated using various Qiskit functions and gates. A list of Qiskit's circuit library can be found in \cite{qiskit_circuit_library}.

\begin{lstlisting}[language=Python,caption={Example of Using the \texttt{QiskitGates.get\_all()} routine from \texttt{randomqc\_qiskit}},label={ex_qiskit},float=*]
>>> from randomqc_qiskit import QiskitGates

>>> qiskit_qcs = QiskitGates.get_all()
>>> len(qiskit_qcs)
102

>> qiskit_qcs.keys()
dict_keys(['C3XGate', 'C3SXGate', 'C4XGate', 'CCXGate', 'DCXGate', 'CHGate', 'CPhaseGate', 'CRXGate', 'CRYGate', 'CRZGate', 'CSGate', 'CSdgGate', 'CSwapGate', 'CSXGate', 'CUGate', 'CU1Gate', 'CU3Gate', 'CXGate', 'CYGate', 'CZGate', 'CCZGate', 'ECRGate', 'HGate', 'IGate', 'PhaseGate', 'RCCXGate', 'RC3XGate', 'RGate', 'RXGate', 'RXXGate', 'RYGate', 'RYYGate', 'RZGate', 'RZZGate', 'XXMinusYYGate', 'XXPlusYYGate', 'SGate', 'SdgGate', 'SwapGate', 'iSwapGate', 'SXGate', 'SXdgGate', 'TGate', 'TdgGate', 'UGate', 'U1Gate', 'U2Gate', 'U3Gate', 'XGate', 'YGate', 'ZGate', 'QFT', 'AND', 'OR', 'XOR', 'InnerProduct', 'NLocal', 'TwoLocal', 'RealAmplitudes', 'EfficientSU2', 'ExcitationPreserving', 'QAOAAnsatz', 'PauliFeatureMap', 'ZFeatureMap', 'ZZFeatureMap', 'StatePreparation', 'Diagonal', 'MCMT', 'MCMTVChain', 'Permutation', 'PermutationGate', 'GMS', 'GR', 'GRX', 'GRY', 'GRZ', 'MCPhaseGate', 'MCXGate', 'MCXGrayCode', 'MCXRecursive', 'MCXVChain', 'RVGate', 'LinearAmplitudeFunction', 'LinearPauliRotations', 'PolynomialPauliRotations', 'PiecewiseLinearPauliRotations', 'DraperQFTAdder', 'CDKMRippleCarryAdder', 'WeightedAdder', 'IntegerComparator', 'QuadraticForm', 'ExactReciprocal', 'FourierChecking', 'GraphState', 'HiddenLinearFunction', 'IQP', 'QuantumVolume', 'PhaseEstimation', 'GroverOperator', 'PhaseOracle', 'EvolvedOperatorAnsatz', 'PauliEvolutionGate'])

>>> qiskit_qcs['ZZFeatureMap']
<qiskit.circuit.quantumcircuit.QuantumCircuit at 0x7f7b81690712>
\end{lstlisting}

\subsection{Benchmarking}
To ensure a systematic and streamlined way to benchmark various quantum transpilers, a dedicated package named \texttt{benchmarq} was developed. At the core of this package is an abstract class named \texttt{TranspilationExperiment}. This abstract class provides a standardized framework for conducting transpilation experiments and evaluating the performance of different quantum transpilers.
The \texttt{TranspilationExperiment} class defines three abstract methods:
\begin{itemize}
\item \texttt{check}: This method checks whether the initial and transpiled circuits are identical, ensuring the correctness of the transpilation process.

\item \texttt{metrics}: This method requires the user to return a dictionary where the keys represent different metrics, and the values are lists of two functions. The first function calculates the metric value for the initial circuit, while the second function calculates the metric value for the transpiled circuit.

\item \texttt{transpile}: This method defines how the transpilation is carried out.
\end{itemize}

The \texttt{TranspilationExperiment} class also includes a \texttt{run} method. This method takes a pandas DataFrame containing the circuits to be used for the transpilation experiment and an iteration count for conducting the experiment multiple times. The run method executes the transpilation experiment, collecting relevant data such as accuracy, speed, and the given metrics. The results are then returned as a pandas DataFrame, ready for analysis.

To illustrate a typical workflow using \texttt{RandomQC} and \texttt{benchmarq}, we will present how to carry two experiments to compare the qBraid-SDK and the pytket extensions abilities to transpile circuits from pytket's circuits to braket's circuits. To this end, we will use four scripts:
\begin{itemize}
    \item \texttt{gen\_circuits.py} (\cref{ex_gen_circuits}): generates the random quantum circuits and stores them in a pandas DataFrame.

    \item \texttt{base.py} (\cref{ex_basepy}): defines a class that inherits from the \texttt{TranspilationExperiment} (here named \texttt{PytketToBraketTranspilationEXP}). This class defines the \texttt{check} and \texttt{metrics} methods.
    
    \item \texttt{bpytket.py} (\cref{ex_bpytket}): This file defines a class that inherits from the class defined in \texttt{base.py}: \texttt{PytketToBraketTranspilationEXP} and implements the transpilation method used by pytket. Additionally, it includes a script to load circuits from the DataFrame output of \texttt{gen\_circuits.py} and runs the experiment using the class defined in this file.
    
    \item \texttt{bqbraid.py}: Similar to the \texttt{bpytket.py} file, this file benchmarks the qBraid-SDK package.
\end{itemize}

\section{Results} \label{sec:res}
The comparison of various quantum transpilers was conducted using the following package versions: \texttt{qbraid=0.5.0}, \texttt{qiskit=0.45.3}, \texttt{pytket=1.24.0}, \texttt{cirq=1.24.0}, \texttt{amazon-braket-sdk=1.69.1}, \texttt{pytket-braket=0.34.1}, \texttt{pytket-cirq=0.34.0}, pytket-\texttt{qiskit=0.48.0}, and \texttt{qiskit\_braket\_provider=0.0.5}. 

A total of 50 circuits were generated for each of \texttt{PureRandom} and \texttt{VQE}  circuit types, and each SDK gate/circuit was randomly generated 50 times. Each transpilation experiment was conducted 5 times.
The evaluation metrics employed include ``correct," which represents the mean number of correctly transpiled circuits with a score of 1 indicating all circuits were correctly transpiled. Moreover,  ``fails" indicates the mean number of times the transpiler threw an error (lower is better); and ``time" represents the mean time taken by the transpiler to transpile the circuit in seconds (lower is better). The ``correct" metric is provided in \cref{tab:table_1} and \cref{tab:table_2} for completeness but has been omitted from subsequent tables as it consistently yielded a score of 1 for all transpilers, i.e: the transpiled circuits were always correct. Indeed, either the transpiler successfully transpiled the circuit without errors, or in case of failures, an error was thrown indicating an inability to transpile.

\Cref{tab:table_1} illustrates the performance comparison of transpilers in terms of correctness, failure rate, and time taken for transpilation across different circuit types. \Cref{tab:table_2} further dissects the comparison based on specific circuit types, namely \texttt{PureRandom}, \texttt{SDKGate}, and \texttt{VQE}. 
The results show that the qiskit-braket-provider exhibits the best performance, with a remarkably low failure rate of only 0.2\%. In contrast, the qBraid-SDK follows with a failure rate of 6.9\%, and pytket extensions rank last with a failure rate of 50.4\%  These results can be attributed to the specialized nature of the qiskit-braket-provider, which is specifically designed for transpilation from Qiskit to Braket. Utilizing a combination of one-to-one transpilation and gate decomposition for unsupported gates via Qiskit's \texttt{transpile} method, the qiskit-braket-provider achieves superior transpilation capabilities.
On the other hand, the qBraid-SDK offers a more generalized approach, enabling transpilation across various quantum SDKs. Its hub-and-spokes architecture centered around Cirq provides respectable transpilation capabilities but at the cost of slower performance compared to the qiskit-braket-provider and the pytket extensions.
Lastly, the pytket extensions rank last in transpilation capabilities due to their one-to-one transpilation scheme, resulting in errors when gates are not present in both the initial and target SDKs. However, this approach allows for a faster transpilation than the qBraid-SDK.

\begin{figure*}[t!]
  \centering
  \includegraphics[width=0.85\textwidth]{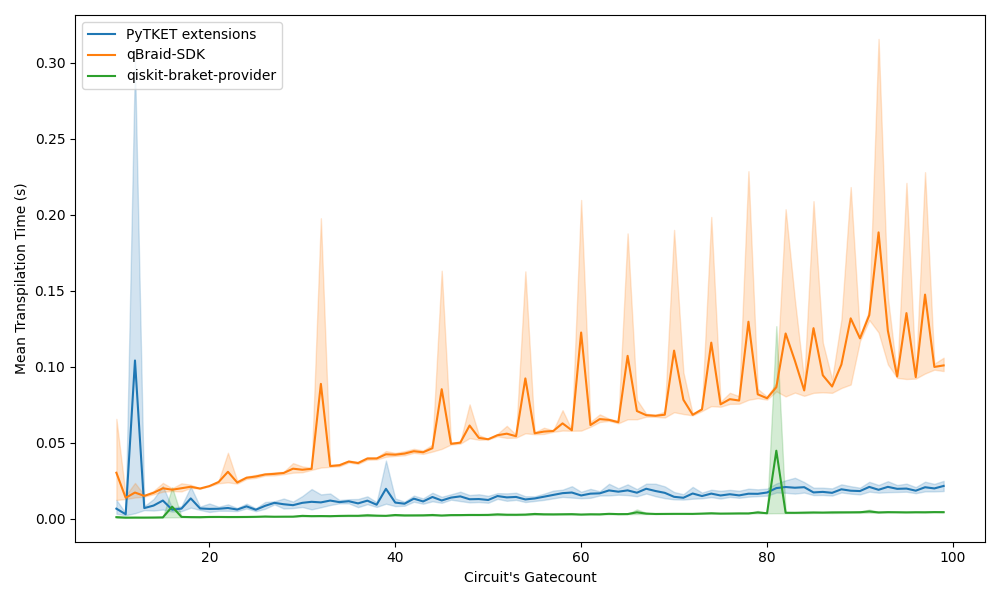}
  \caption{Mean transpilation time (from qiskit to braket) vs Circuit's Gatecount comparison between the qiskit-braket-provider, the qBraid-SDK, and the pytket extensions.}
  \label{fig:time_scale_1}
\end{figure*}

\begin{figure*}[t!]
  \centering
  \includegraphics[width=0.85\textwidth]{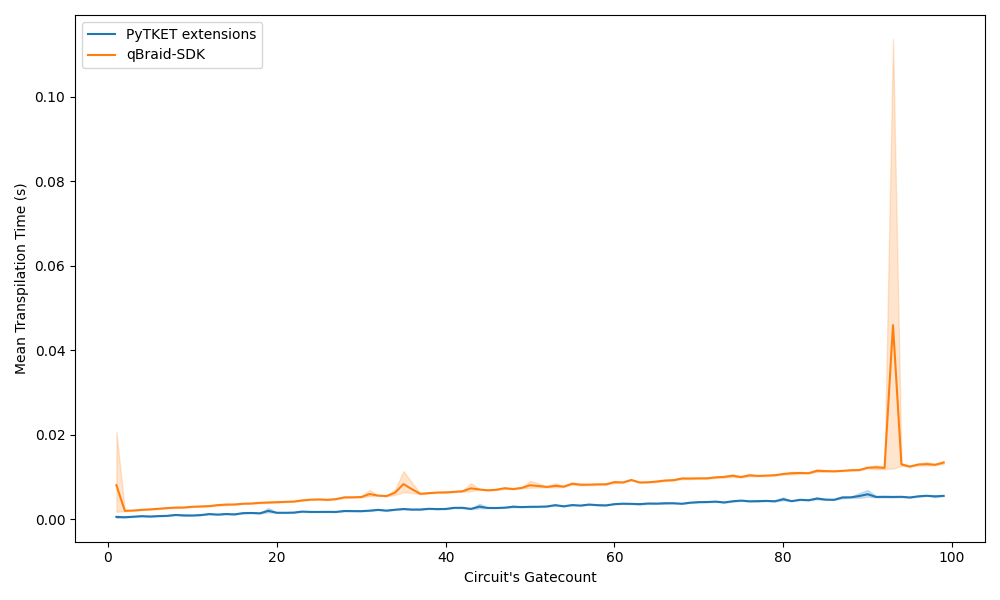}
  \caption{Mean transpilation time (from braket to cirq) vs Circuit's Gatecount comparison between the qBraid-SDK and the pytket extensions.}
  \label{fig:time_scale_2}
\end{figure*}

Although, the pytket extensions appears to be the fastest transpilers based on the results of Tables \ref{tab:table_1}, \ref{tab:table_2}, and \ref{tab:table_3}, this observation requires further examination. The time column in those tables reflects the average time for successfully transpiled circuits. However, the pytket extensions exhibits a significant failure rate, particularly when transpiling more complex circuits. To conduct a fair comparison, two additional experiments were conducted in which the \texttt{range\_gates} method of RandomQC was employed to generate circuits ranging from 10 to 100 qubits. These circuits were selected to be transpilable by the qiskit-braket-provider, the qBraid-SDK, and the pytket extensions, ensuring a consistent basis for comparison. In the first experiment, the transpilation was carried from qiskit to braket. In the second experiment, the transpilation was carried from braket to cirq. The results are shown in \cref{fig:time_scale_1} and \cref{fig:time_scale_2} for the first and second experiments respectively. Both these figures show the mean transpilation time as the circuit's gatecount increases. The qiskit-braket-provider demonstrates the best performance, with the lowest transpilation times observed as the gatecount increases. The pytket extensions follow closely behind, while, the qBraid-SDK lags behind both.

\section{Conclusion} \label{sec:conc}
In conclusion, our comprehensive evaluation of various quantum transpilers provides valuable insights into their capabilities and performances. Across different circuit types and transpilation tasks, we observed distinct strengths and weaknesses among the evaluated transpilers. For transpilation from Braket to Qiskit, the dedicated qiskit-braket-provider package emerges as the top performer, excelling in both capabilities and speed. In contrast, for transpilation from Pytket to Qiskit, the pytket extension (\texttt{pytket-qiskit}) outperforms the qBraid-SDK package in terms of both capabilities and speed. For other transpilation scenarios, the qBraid-SDK demonstrates the best overall capabilities and should be considered the go-to solution. Its hub-and-spokes architecture centered around Cirq provides robust transpilation capabilities, albeit at a slightly slower speed compared to specialized solutions like the qiskit-braket-provider and to the one-to-one transpilation scheme of the pytket extensions.

In terms of development recommendations, the approach adopted by the qiskit-braket-provider, which combines one-to-one transpilation with gate decomposition, appears to be the most promising. This method not only ensures high transpilation capabilities but also enhances speed, as evidenced by its top ranking in both capabilities and speed. We thus recommend adopting similar strategies in the development of quantum transpilers to improve both their capabilities and efficiency. By doing so, advancements in the field of quantum computing can be accelerated, and the seamless integration of diverse hardware and software platforms can be facilitated.

\onecolumngrid

\begin{lstlisting}[language=Python,caption={Random Quantum Circuit Generation Script (\texttt{gen\_circuits.py})},float=*b,label={ex_gen_circuits}]
import pandas as pd
from randomqc import PureRandom
from randomqc.variational import VQE
from randomqc_pytket import PyTKETGates

nb_circuits = 50
nb_gates_runs = 50

circuits = []

# Generating pure random quantum circuits
for i in range(nb_circuits):
    circ = PureRandom()
    circ_sdk = circ.to('pytket')
    circuits.append({
        'circuit_name': circ.name, 
        'nb_qubits': circ.nb_qubits,
        'circuit': circ_sdk,
        'qasm2': circ.to('qasm2'),
    })

# Generating PyTKET gates
for i in range(nb_gates_runs):
    gates = PyTKETGates.get_all()
    for name, circ in gates.items():
        circuits.append({
            'circuit_name': f'PyTKET-{name}', 
            'nb_qubits': circ.n_qubits,
            'circuit': circ,
            'qasm2': '[PYTKET]',
        })

# Generating VQE-type random circuits
for i in range(nb_circuits):
    circ = VQE()
    circ_sdk = circ.to('pytket')
    circuits.append({
        'circuit_name': circ.name, 
        'nb_qubits': circ.nb_qubits,
        'circuit': circ_sdk,
        'qasm2': circ.to('qasm2'),
    })

df = pd.DataFrame(circuits)
df.to_pickle('pytket.pkl')
\end{lstlisting}

\begin{lstlisting}[language=Python,caption={Base Transipilation Experiment Script (\texttt{base.py})},label={ex_basepy},float=*]
from benchmarq.transpilation import TranspilationExperiment

# Utility function that check that two unitaries (np.array) are equal up to global phase
from benchmarq.transpilation import assert_allclose_up_to_global_phase

# Utility functions that transform an SDK circuits to a unitary (np.array) 
from benchmarq.transpilation.sdk_utility import pytket_to_unitary, braket_to_unitary 

# Utility functions that calculates the Depth and Gatecount for Pytket and Braket circuits respectively
from benchmarq.transpilation.sdk_utility import pytket_depth, pytket_gatecount, braket_depth, braket_gatecount 
    
class PytketToBraketTranspilationEXP(TranspilationExperiment):
    def check(self, circuit_pytket, circuit_braket):
        try:
           assert_allclose_up_to_global_phase(pytket_to_unitary(circuit_pytket), 
                                              braket_to_unitary(circuit_braket), 
                                              atol=1e-7)
           return 1

        except Exception as e:
            print(e)
            return 0
        
    def metrics(self):
        return {
            'depth': [pytket_depth, braket_depth],
            'gatecount': [pytket_gatecount, braket_gatecount]
        }
\end{lstlisting}

\begin{lstlisting}[language=Python,caption={Pytket Transpilation Implementation (\texttt{bpytket.py})},label={ex_bpytket},float=*]
import pandas as pd

from pytket.extensions.qiskit import qiskit_to_tk
from pytket.extensions.braket import tk_to_braket


from base import PytketToBraketTranspilationEXP

class UsingPyTKET(PytketToBraketTranspilationEXP):
def __init__(self):
    self.name = 'pytket-braket==0.34.1'
    
def transpile(self, circuit):
    circ, _, _ =  tk_to_braket(circuit)
    return circ

if __name__ == "__main__":
    initial_circuits = pd.read_pickle('pytket.pkl')
    dftket = UsingPyTKET().run(circuits=initial_circuits, iter=5)
    dftket.to_csv('pytket.csv')
\end{lstlisting}

\begin{table*}[]
\centering
\caption{Overview Comparison of Quantum Transpilers (average over all tasks)}
\label{tab:table_1}
\begin{tcolorbox}[tab2,tabularx={X|YYY},boxsep=2mm]
     & Correct & Fails & Time (s) \\
Transpiler &  &  &  \\\hline
pytket extensions & 1.000000 & 0.504258 & 0.001160 \\
qBraid-SDK & 1.000000 & 0.068935 & 0.023555 \\
qiskit-braket-provider & 1.000000 & 0.002692 & 0.008127 \\
\end{tcolorbox}
\end{table*}

\begin{table*}[]
\centering
\caption{Overview Comparison of Quantum Transpilers for Different Circuit Types}
\label{tab:table_2}
\begin{tcolorbox}[tab2,tabularx={X>{\hsize=1.4\hsize}X|YYYY},boxsep=2mm]
 &  & Correct & Fails & Time (s) \\
Circuit type & Transpiler &  &  &  \\
\hline
\multirow[t]{3}{*}{PureRandom} & pytket extensions & 1.000000 & 0.663333 & 0.015469 \\
 & qiskit-braket-provider & 1.000000 & 0.000000 & 0.024760 \\
 & qBraid-SDK & 1.000000 & 0.000000 & 0.086634 \\
\hline
\multirow[t]{3}{*}{SDKGate} & pytket extensions & 1.000000 & 0.504813 & 0.000845 \\
 & qiskit-braket-provider & 1.000000 & 0.002745 & 0.007989 \\
 & qBraid-SDK & 1.000000 & 0.070410 & 0.022340 \\
\hline
\multirow[t]{3}{*}{VQE} & pytket extensions & 1.000000 & 0.293333 & 0.014963 \\
 & qiskit-braket-provider & 1.000000 & 0.000000 & 0.005484 \\
 & qBraid-SDK & 1.000000 & 0.000000 & 0.066100 \\
\end{tcolorbox}
\end{table*}

\begin{table*}[]
\centering
\caption{Overview Comparison of Quantum Transpilers for Various SDK Pairs}
\label{tab:table_3}
\begin{tcolorbox}[tab2,tabularx={XX>{\hsize=1.4\hsize}X|YYY},boxsep=2mm]
 &  &  & Fails & Time (s) \\
From & To & Transpiler &  &  \\
\midrule
\multirow[t]{6}{*}{braket} & \multirow[t]{2}{*}{cirq} & pytket extensions & 0.432432 & 0.000324 \\
 &  & qBraid-SDK & 0.006486 & 0.002303 \\
\cline{2-5}
 & \multirow[t]{2}{*}{pytket} & pytket extensions & 0.270270 & 0.000194 \\
 &  & qBraid-SDK & 0.006486 & 0.015800 \\
\cline{2-5}
 & \multirow[t]{2}{*}{qiskit} & pytket extensions & 0.270270 & 0.000895 \\
 &  & qBraid-SDK & 0.006486 & 0.013475 \\
\cline{1-5} \cline{2-5}
\multirow[t]{6}{*}{pytket} & \multirow[t]{2}{*}{braket} & pytket extensions & 0.479600 & 0.000557 \\
 &  & qBraid-SDK & 0.320000 & 0.017467 \\
\cline{2-5}
 & \multirow[t]{2}{*}{cirq} & pytket extensions & 0.517600 & 0.000235 \\
 &  & qBraid-SDK & 0.320000 & 0.015564 \\
\cline{2-5}
 & \multirow[t]{2}{*}{qiskit} & pytket extensions & 0.000000 & 0.001762 \\
 &  & qBraid-SDK & 0.060000 & 0.007765 \\
\cline{1-5} \cline{2-5}
\multirow[t]{7}{*}{qiskit} & \multirow[t]{3}{*}{braket} & pytket extensions & 0.716731 & 0.002726 \\
 &  & qiskit-braket-provider & 0.002692 & 0.008127 \\
 &  & qBraid-SDK & 0.013269 & 0.046340 \\
\cline{2-5}
 & \multirow[t]{2}{*}{cirq} & pytket extensions & 0.783846 & 0.001261 \\
 &  & qBraid-SDK & 0.013269 & 0.029017 \\
\cline{2-5}
 & \multirow[t]{2}{*}{pytket} & pytket extensions & 0.452115 & 0.001340 \\
 &  & qBraid-SDK & 0.009808 & 0.021215 \\
\end{tcolorbox}
\end{table*}

\clearpage
\section*{Acknowledgments}

The current work was carried out during a remote internship at qBraid. We would like to express our gratitude to the qBraid team for their financial and technical support, guidance, and valuable insights throughout the duration of this research. Their expertise and resources significantly contributed to the success of this study. This document has been produced with the financial assistance of the European Union (Grant no. DCI-PANAF/2020/420-028), through the African Research Initiative for Scientific Excellence (ARISE), pilot programme. ARISE is implemented by the African Academy of Sciences with support from the European Commission and the African Union Commission. The contents of this document are the sole responsibility of the author(s) and can under no circumstances be regarded as reflecting the position of the European Union, the African Academy of Sciences, and the African Union Commission. We are grateful to the Algerian Ministry of Higher Education and Scientific Research and DGRST for the financial support.


\bibliography{bibliography.bib}
\end{document}